\documentclass[sigconf]{acmart}

\AtBeginDocument{%
	}

\setcopyright{acmlicensed}
\copyrightyear{2026}
\acmYear{2026}
\acmDOI{XXXXXXX.XXXXXXX}
\acmConference[Preprint]{Make sure to enter the correct
	conference title from your rights confirmation email}{2026}{}
\acmISBN{978-1-4503-XXXX-X/2018/06}

\usepackage{hyperref}
\usepackage{url}  
\usepackage{xcolor}

\definecolor{quoteClr}{RGB}{ 10, 60, 130} 

\usepackage{tikz}

\usepackage{dirtytalk}

\usepackage{listings}

\usetikzlibrary {positioning}
\usetikzlibrary{shapes.geometric, arrows, backgrounds, fit, calc}

\pgfdeclarelayer{background}
\pgfdeclarelayer{foreground}
\pgfsetlayers{background,main,foreground}

\newcounter{exmpl}
\newenvironment{exmpl}[1][]{\medskip\refstepcounter{exmpl}\par
	\noindent $\blacktriangleright$ \textbf{Example \theexmpl:~ #1} \\ \noindent}{\hfill $\blacktriangleleft$ \par  \smallskip\noindent%
	\ignorespacesafterend \normalsize 
	}

\newcommand{\DpartnerA}[2]{\noindent \textbf{#1:}~\textcolor{magenta}{\say{#2}}
	
	\smallskip}

\newcommand{\DpartnerB}[2]{\noindent \textbf{#1:}~\textcolor{blue}{\say{#2}}
	
	\smallskip}
	
\newcommand{\DpartnerAC}[2]{\noindent \textbf{#1:}~\textcolor{magenta}{\say{#2}}}

\newcommand{\DpartnerBC}[2]{\noindent \textbf{#1:}~\textcolor{blue}{\say{#2}}}

\newcommand{\DDescription}[1]{\textless *#1*\textgreater}

\newcommand{\CodeFormat}[1]{\mbox{\textit{\textless#1\textgreater}}}

\newcommand{\DCode}[1]{\hfill \CodeFormat{#1}
	
	\smallskip}
	
\newcommand{\DCodeB}[1]{\hfill \CodeFormat{#1}}
	
\newcommand{\DCodeTwo}[2]{\hfill \CodeFormat{#1} \CodeFormat{#2}
	
	\smallskip}

\newcommand{\DCodeThree}[3]{\hfill \CodeFormat{#1} \CodeFormat{#2} \CodeFormat{#3}
	
	\smallskip}

\newcounter{K}
\renewcommand{\theK}{(\arabic{K})}
\newcommand{\instK}[1]{\refstepcounter{K}\label{#1}\theK}
\newcommand{\Elem}[2]{\renewcommand{\theK}{#1{\arabic{K}}}\instK{#2}}

\newcommand{\newQ}[1]{\Elem{Q}{#1}}

\newcommand{\Qby}[3]{%
	\textcolor{quoteClr}{\emph{\say{#3}} (\newQ{#2}:#1)}%
}

\usepackage{tabularx,ragged2e}
\usepackage{makecell} 
\newcolumntype{C}{>{\arraybackslash}X} 
\newcolumntype{Q}{>{\Centering\arraybackslash}X} 

\tikzstyle{SuccessFactor} = [circle, minimum width=1.5cm, text width=1.5cm, text centered, inner sep=1pt, draw=black, fill=green!30, font=\scriptsize]
\tikzstyle{RiskFactor} = [circle, minimum width=1.cm, text width=1.5cm, text centered, inner sep=1pt, draw=black, fill=red!30, font=\scriptsize]
\tikzstyle{Task} = [rectangle, minimum width=1.8cm, minimum height=1cm, text width=1.7cm, inner sep=1pt, text centered, draw=black, fill=gray!10, font=\scriptsize]
\tikzstyle{SubTask} = [rectangle, minimum width=1.6cm, minimum height=1cm, text width=1.5cm, inner sep=1pt, text centered, draw=black, fill=gray!5, font=\scriptsize, dashed]
\tikzstyle{BadTask} = [rectangle, minimum width=1.8cm, minimum height=1cm, text width=1.7cm, inner sep=1pt, text centered, draw=black, fill=red!10, font=\scriptsize]
\tikzstyle{BadSubTask} = [rectangle, minimum width=1.6cm, minimum height=1cm, text width=1.5cm, inner sep=1pt, text centered, draw=black, fill=red!5, font=\scriptsize, dashed]
\tikzstyle{Legend} = [rectangle, minimum width=2cm, minimum height=1.5cm, text width=2cm, inner sep=4pt, text centered, draw=black, font=\scriptsize]
\tikzstyle{posArrowL} = [green!50!black!50, , thick, line width=1mm,->,>=stealth]
\tikzstyle{posArrowS} = [green!50!black!50, thick, line width=0.66mm,->,>=stealth]
\tikzstyle{negArrowL} = [red!50!black!60, thick, line width=1mm,->,>=stealth]
\tikzstyle{negArrowS} = [red!50!black!60, thick, line width=0.66mm,->,>=stealth]

\tikzstyle{subConceptA} = [o-stealth, black, thick, dashed]

\usetikzlibrary{patterns}

\newcommand{\basicTimeline}[1]{
	\newcounter{recHour}
	\setcounter{recHour}{0}
	\newcount\minuteOne; \minuteOne=0 
	\def\w{0.8\columnwidth}    
	\def\n{#1}     
	\def\lt{0.40} 
	\def\lf{0.20} 
	\def\lo{0.0}  

	\draw[->,thick] (-\w*0.03,0) -- (\w*1.03,0);
	
	\foreach \tick in {0,1,...,\n}{
		\def\x{{\tick*\w/\n}}
		\def\minute{\the\numexpr \minuteOne+\tick*10-\value{recHour}*60 \relax}
		\ifnum\minute > 59
		\stepcounter{recHour}
		\def\minute{\the\numexpr \minuteOne+\tick*10-\value{recHour}*60 \relax}
		\fi
		\draw[thick] (\x,\lt) -- (\x,-\lt) 
		node[below] {
			\ifnum\minute<10
			{\footnotesize \the\value{recHour}:0\minute h}
			\else
			{\footnotesize \the\value{recHour}:\minute h}
			\fi
		};
		
		\ifnum \tick<\n
		\draw[thick] ({(\x+\w/\n/2)},0) -- ({(\x+\w/\n/2)},\lf); 
		\foreach \ticko in {1,2,3,4,6,7,8,9}{
			\def\xo{{(\x+\ticko*\w/\n/10)}}
			\draw[thick] (\xo,0) -- (\xo,\lo);  
		}\fi
	}
}

\def\minuteLabel(#1,#2){
	\node[above] at ({(#1-\minuteOne)*\w/\n/10},\lt) {#2};}

\def\minuteArrowLabel(#1,#2,#3,#4){ 
	\def\xy{{(#1-\minuteOne)*\w/\n/10}}; \pgfmathparse{int(#2*100)};
	\ifnum \pgfmathresult<0
	\def\yyp{{(\lt*(0.90+#2))}}; \def\yyw{{(\yyp-\lt*#3)}}
	\draw[<-,thick,black,align=center] (\xy,\yyp) -- (\xy,\yyw) node[below,black] at (\xy,\yyw) {#4};
	\else
	\def\yyp{{(\lt*(0.10+#2)}}; \def\yyw{{(\yyp+\lt*#3)}}
	\draw[<-,thick,black,align=center] (\xy,\yyp) -- (\xy,\yyw) node[above,black] at (\xy,\yyw) {#4};
	\fi}

\newcommand{\timelineExample}[5]{
	\draw[#3, pattern={north east lines}, pattern color = #3] (#1*\w/\n/10,0) rectangle (#2*\w/\n/10,0.5) node[above = #5+0.2cm, xshift=0.8cm, text width=2cm] {\footnotesize #4};
}

\begin{document}
\title[Qualitative Analysis of the Teacher and Student Roles in Pair Programming]{Qualitative Analysis of the Teacher and Student Roles \\ in Pair Programming}

\author{Linus Ververs}
\email{linus.ververs@fu-berlin.de}
\affiliation{%
	\institution{Freie Universität Berlin}
	\country{Germany}
}

\author{Trang Linh Lam}
\email{lamt02@zedat.fu-berlin.de}
\affiliation{%
	\institution{Freie Universität Berlin}
	\country{Germany}
}

\author{Lutz Prechelt}
\email{prechelt@inf.fu-berlin.de}
\affiliation{%
	\institution{Freie Universität Berlin}
	\country{Germany}
}

\renewcommand{\shortauthors}{Ververs, Lam \& Prechelt}

\begin{abstract}
	\emph{Background:} Pair programming is a well-established and versatile
	agile practice.
	Previous research has found it to involve far more different roles than
	the well-known Driver and Observer/Navigator roles.
	Pair programming often involves heavy knowledge transfer from mainly
	one partner to the other.
	\\
	\emph{Objective:} Understand how to fill the ensuing Teacher and Student 
	roles well (positive behavioral patterns). 
	Understand how they may break (anti-patterns).
	\\  
	\emph{Method:}
	Open coding and axial coding of 17 recorded pair programming sessions 
	with 18 developers from 5 German software companies, 
	plus interviews with 6 different developers from 4 other German companies.\\
	\emph{Results:} We describe six facets of effective Teacher behavior 
	(e.g. Prioritizing Knowledge Transfer) and
	two facets of effective Student behavior
	(e.g. Expressing Knowledge Wants).
	We describe four harmful would-be-Teacher behaviors
	(e.g. Pushing Unwanted Knowledge), 
	and
	one harmful would-be-Student behavior
	(Failing to Provide a Back Channel).
	\\
	\emph{Conclusions:}
	The role facets can serve as learning goals and to-do list for developers
	who want to develop strong pair programming skill.
	The anti-patterns can serve as warnings for one's own general behavior
	and as triggers for immediate meta-discussion if they occur within
	a pairing session.
\end{abstract}




\maketitle


\section{Introduction}

In the most-cited definition of pair programming, \say{two programmers jointly produce one artifact (design, algorithm, code)} \cite{WilKesCun00}. 
To which the second-most-cited definition adds 
\say{It isn’t one person programming while another person watches. [\ldots] 
	Pair programming is a dialog between two people trying to simultaneously program (and analyze and design and test) and understand together how to program better.} \cite{Beck99}.
Zieris differentiates pair programming as a work mode, to be used as needed, 
from pair programming as a practice, to be used constantly \cite[Section 2.3.1]{Zie20}.
We are talking about the work mode here.

Pair programming is claimed to have many advantages 
(see for example \cite[Chapter 3]{WilKes03}) 
that can be categorized into three dimensions: 
knowledge transfer between participating developers \cite{PloShaLin15,Zie20,ZiePre20,VanLas05}, 
increased quality through the four-eyes principle with potentially better decisions
\cite{HanDybAri09, FuGraBro17, VanLas05, Zac11, AriGalDyb07}, and 
increased speed through more ideas being generated (especially during debugging), 
by reducing handover times, and by increasing focus \cite{HanDybAri09, SilSucVla12}.

Previous pair programming research has suggested the idea of a
pair-programming-specific skill that is distinct from the technical
programming skill~\cite{ZiePre14,ZiePre21}.
We share the perspective that empirical software engineering research
should produce insights that are actionable~\cite{SjoDybAnd08}.
We aim at understanding
further elements of pair-programming skill and formulate it in the form
of advice to practitioners.

We work with recordings of actual industrial pair-programming sessions.
In several of these, we noticed they did not appear to go well
in terms of the pair collaboration,
so we decided to understand which elements of pair-programming skill
were lacking in these cases.
We quickly found many of the cases to involve teacher/student constellations,
where there was a large gap in the knowledge between the pair members.
Previous work had already suggested a framework for describing
roles in pair-programming \cite{SalPreZie13}.
A role description is somewhat similar to a design pattern \cite{GamHelJoh95}
(with a unique name, context description, elements, variants, and discussion)
and is therefore a good format for advice to practitioners.

So we decided to describe two roles in this manner: 
\emph{Teacher} and \emph{Student}.
It turned out we could find constructive elements of behavior for filling
each role, as well as problematic elements.

Our research eventually led to two articles, both currently under
review: The present one and another that 
develops a grounded theory explaining an important category of dysfunctional
behavior~\cite{NN26b-CHASE}.

\subsection{Research Questions}

Our research uses the attitude and elements of GTM,
Grounded Theory Methodology \cite{StrCor90, Charmaz14},
which works from a research interest, not research questions.
The research interest was understanding the \emph{Teacher} and \emph{Student} roles.
For the benefit of readers less familiar with GTM,
we additionally formulate four explicit research questions:
\begin{itemize}
  \item RQ Tpos: Formulate the role of an effective \emph{Teacher} in pair programming.
  \item RQ Spos: Formulate the role of an effective \emph{Student} in pair programming.
  \item RQ Tneg: Formulate behaviors that tend to break the \emph{Teacher} role.  \item RQ Sneg: Formulate behaviors that tend to break the \emph{Student} role.
\end{itemize}

\subsection{Research Contributions}

We explain the roles of \emph{Teacher} and \emph{Student} 
as well as problematic behaviors in large-knowledge-gap situations
in a manner that allows practitioners to relate them to their own behavior
as pair members and to improve it if necessary.

\subsection{Article Overview}

We describe related work
(Section~\ref{SecRelatedWork}),
our data
(Section~\ref{SecPPSessions}),
our coding process 
(Section~\ref{SecCoding}),
and our concept definitions writing style
(Section~\ref{sec:conceptdefs}),
then define the \emph{Teacher} role
(Section~\ref{ResDefTeacherRole})
and \emph{Student} role 
(Section~\ref{ResStudentRole}) and
explain anti-patterns when taking up either role
(Sections~\ref{Res_Fail_T} and~\ref{Res_Fail_S}).
We discuss limitations
(Section~\ref{sec_lim}),
and conclude
(Section~\ref{sec_concl}).

\section{Related Work}\label{SecRelatedWork}

Salinger et al.'s pair programming roles framework describes how 
pair partners fulfill many more different, conceptually separate functions 
than those described for the common and hopelessly overloaded driver and observer 
(or navigator) roles \cite{SalPreZie13}. 
They define a meta-model for defining roles in pair programming. 
A role is defined by facets, which are observable actions taken by one developer during a session. 
A role can be defined by multiple facets, but \say{[t]here is no
	fixed rule as to which or how many facets must be observed to conclude the respective pair member has assumed the Role} \cite{SalPreZie13}. 
We built upon this framework for role definition. 

Salinger and Prechelt created the \say{Base Layer} which is a collection of concepts that serve as 
\say{a foundation for qualitative research into pair programming [...] that aims at explaining the pair programming	process} \cite{SalPre13}. 
These concepts stem from analyzing recorded industrial pair programming sessions and focus on the dialog of a pair. These concepts apply to our data well and our analysis makes use of them.

Zieris, as part of his qualitative analysis of knowledge transfer in
pair programming introduced the concepts of \emph{Togetherness}~\cite[Chapter 6]{Zie20} and of
\emph{Push} and \emph{Pull} modes of knowledge transfer
\cite[Chapter 9]{Zie20}.
We encountered these phenomena and others also previously
introduced by Zieris \cite{ZiePre14,Zie20} and used the respective
concepts internally in our own coding (see Section \ref{sec:methods}).

Zieris and Prechelt study knowledge transfer in pair programming
\cite{ZiePre14,ZiePre16,ZiePre21}. 
They work with the same data set \cite{ziepre20-pp-ind} as we do here.
In \cite{ZiePre21}, they discuss one example from session PA3
that we also discuss (Example~\ref{PA3_MagicNumber}). 
They use it to illustrate the anti-pattern \say{Drowning the Partner} 
that is characterized by \say{too many explanations that 
	(a) go far beyond the task and are hence not expedient and 
	(b) also threaten the pair’s \emph{Togetherness}}.

Chong and Hurlbutt conducted an ethnographic observation of pair
programming \cite{ChoHur07}.
They comment on roles and on knowledge
disbalances (important topics for us): \say{[a]side from the task of
  typing, we found no consistent division of labor between the
  \say{driver} and the \say{navigator}} and \say{[t]he gaps in
  expertise between the programmers on Team B clearly influenced pair
  programming interactions on the team. On Team B, the member of the
  pair with greater expertise drove the bulk of the programming
  discussions}.

Begel and Nagappan ran a survey at Microsoft \cite{BegNag08}. 
They focused on benefits and problems of pair programming. 
They list \say{skill differences} (which play a role in our analysis)
among the top problems.

There are lots of pair programming research using other kinds of data
(from educational contexts)
or research methods (such as quantitative surveys and controlled experiments).

Some of these describe phenomena that, once one has seen our results,
can be recognized as instances of similar phenomena like those  we discuss:
the Teacher role is difficult \cite{BowJarCul19},
a one-sided knowledge gap can lead to problems \cite{BowJarCul19,DomColHev03},
pairs need to learn pair programming to become fully productive \cite{VanLas05}.

Based on controlled experiments, Hannay et al. conclude that pair
programming leads to a small positive effect on quality, a medium
positive effect on duration, and a negative effect on effort
\cite{HanDybAri09}.

\section{Methods and Data} \label{sec:methods}

We use elements of Grounded Theory Methodology (GTM) mostly based on
Strauss \& Corbin's interpretation~\cite{StrCor90}, but also
incorporate Charmaz's constructivist perspective and warnings
\cite{Charmaz06,Charmaz14}.

We do not apply the full set of GTM practices and do not claim the present work
to be a GTM contribution, because neither is required for the research goal
we aim for. 
However, we do use open coding, axial coding, constant comparison, memoing,
and theoretical sampling
and apply a theoretical coding style for our concepts.

For our research, we rely on a preexisting data set of approximately 100 hours of recorded pair programming sessions from 13 companies which were recorded between 2006 and 2018, the PP-ind repository \cite{ziepre20-pp-ind}. 
57 developers participated in these recordings. 
They work on their respective company's software and \emph{real} tasks.
The videos we analyzed consist of a screen recording of the IDE and video with audio of the developers. 
For validation, these data were complemented by six 30-to-60 minute semi-structured
interviews with six additional developers from four different companies.
We used these data in the analysis mostly internally, but will also
refer to them in the article a few times.
The ethics regime of our university does not require ERB approval for
such work (our interview partners are peers).
Transcripts of these interviews (in German) are available on
Figshare\footnote{\url{https://figshare.com/s/24b1e48102189edd6853}}.
The video data make visible many proprietary details of the companies' source
code; they also allow identifying developers.
Therefore, they cannot be published.

Our qualitative research process started when we noticed how in one of the recorded sessions the pair dynamic seemed to be \say{off} in the context
of knowledge transfer. 
Normally, knowledge transfer is viewed as a \emph{positive} outcome of pair programming, 
so we decided to investigate this further.

\subsection{Session Description} \label{SecPPSessions}
Our analysis is based on 17 sessions which we
name according to the naming scheme of the PP-ind repository~\cite{ziepre20-pp-ind}, 
where the first letter represents the company, the second letter represents the technical context, i.e., 
the software system the pair is working on, 
and the number represents the chronological order of pair programming sessions 
within that technical context. 
For example, session PA3 and PA4 are pair programming sessions that were recorded in company P. In both sessions, the developers worked on the same context A and session PA3 happened before PA4. 

Developers are named after the company and the chronological order of their first appearance in the 
sessions. 
For example P1 is the first developer of company P that appears in the data.

We analyzed sessions AA1, CA1, CA2, CA5, DA2, DA4, DA5, DA6, JA1-JA5, and PA1-PA4. 
We will present and discuss episodes taken from CA1, CA5, DA2, PA3, and PA4.

\subsubsection{PA3} \label{sessionPA3}
Session PA3 is 1:30 hours long and is the above-mentioned starting point of our research. 

``Developers P1 (who is more knowledgeable in the backend) and P3 (more frontend)
continue the implementation of a new API endpoint, which P3 already started. 
In PA3, P3 shows his existing implementation for which they [then] write tests; 
P1 explains backend-related software development best practices'' \cite{ziepre20-pp-ind}.

\begin{figure}[htp]
	\begin{tikzpicture}[]
		\basicTimeline{9.1}
		
		\timelineExample{2.6}{3}{blue}{\ref{PA3_Tab}}{0}
		\timelineExample{35.1}{36.75}{blue}{\ref{PA3_expectedArgument} \ref{PA3_assertSame2sub}}{0}
		\timelineExample{29}{31.6}{red}{\ref{PA3_MagicNumber}}{0}
		\timelineExample{11}{12.5}{red}{\ref{PA3_AssertSame}}{0.1}
		\timelineExample{53}{55}{red}{\ref{PA3_NullObject}}{0}
		\timelineExample{14}{16}{red}{\ref{PA3_runTests}}{8}
		\timelineExample{4.1}{4.5}{red}{\ref{PA3_CodeStyle}}{8}
	\end{tikzpicture}
	\caption{Timeline of session PA3, where each example we discuss as part of this article is annotated. We highlighted the problematic episodes with red and the less problematic ones in blue.}
	\label{Timeline_PA3}
\end{figure}
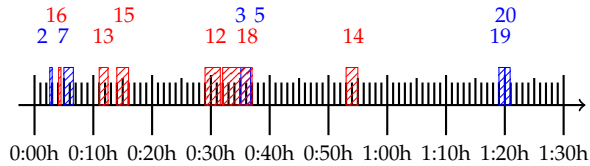

Most examples we will discuss here are taken from this session. 
Figure \ref{Timeline_PA3} shows the chronological ordering and location
of these examples within session PA3.

After session PA3, the researcher who was recording the session overheard a conversation between P1 and P3 in the kitchen. His notes state that P3 criticizes that \say{P1 often gives unsolicited lectures.} 
P1 explains his behavior by saying that \say{there are topics where the other person does not even realize
	 that they should ask a question.} 
At the end of their exchange, P3 acknowledges \say{that many small and large explanations were helpful. 
	That's why he does not want to switch them off completely.}\footnote{Citations translated from the researcher's German field notes.}

\subsubsection{PA4} \label{sessionPA4}
PA4 happens the day after PA3 and is 1:28 hours long. 
P1 and P3 continue ``their work and implement the database access which causes them problems 
because of some idiosyncrasy of their object-relational (OR) mapper'' \cite{ziepre20-pp-ind}.

PA4 is interesting because it features the same pair constellation as PA3, but with a twist. 
As described above, P1 and P3 discussed problems in their pair
programming process in between the sessions (meta-communication).
We can compare behavior between both sessions, to see effects of this
discussion. 

\subsubsection{CA1 and DA2}
These two sessions are characterized by a crucial knowledge gap
between the partners~\cite[Chapter 11]{Zie20}
and were chosen as part of a theoretical sampling step. 

In CA1, the pair works on the GUI of a geographic information system. 
C1 started to work on the task before the session, but C2 possesses more relevant knowledge,
so he quickly takes the lead. The session is 1:18 hours long. 

In DA2 the pair (from a different company) works on a new toolbar as part of the GUI of a
customer-relationship system. 
Although D4 is newly hired by company D, he has more knowledge of Java than is partner D3. 
In the beginning, D4 asks a lot of questions about the software system and the task 
but later he takes the lead and D3 falls behind. The session is 2:26 hours long.

\subsubsection{CA5}
CA5 was chosen in a different theoretical sampling step:
Here, the partners C3 and C4 work together very harmoniously. 
We analyzed CA5 as a contrasting example to behaviors we had analyzed previously. 
In this session, the pair starts work on a new feature for their geographic information system. 
The session is 1:25 hours long.

\subsection{Open Coding \& Code Development} \label{SecCoding}

Here, we explain how the coding process evolved and which concepts emerged when.

\subsubsection{First Generation of Codes: Capturing the Problematic Behavior}

When we started looking at PA3, we quickly found the collaboration appeared dysfunctional. 
We therefore started our open coding with the goal to describe and explain what we felt was wrong and why.
We noticed that most of the dysfunctional behavior of the pair had to do with 
knowledge transfer from P1 to P3. 
This is when we decided, after we had defined our first codes, to build upon 
previous qualitative research on pair programming and concepts defined there. 
We used Salinger \& Prechelt's Base Layer~\cite{SalPre13} to structure the dialog of the pairs 
and to highlight knowledge transfer activities. 
These codes take a verb\_object form, e.g., propose\_step.
Here are some groups of concepts we applied:
\begin{itemize}
	\item \{ask/propose/agree/challenge/disagree/amend/ decide\}\_
      \{design/strategy/step\}:
      for decision making regarding the design of the product (design)
      or the process (strategy/step). 
	\item \{explain/agree/challenge/disagree/amend\}\_finding: to
      verbalize new insights 
	\item \{propose/agree/challenge/disagree/amend\}\_hypothesis: to discuss other hypotheses or conjectures  
	\item \{ask/explain/agree/challenge/disagree\}\_knowledge: for remaining cases of knowledge transfer activities 
	\item \{explain/ask\}\_standard of knowledge: for exchanges
      regarding what the partners know or do not know.
\end{itemize}
Refer to \cite{SalPre13} for the full (and highly detailed) concept definitions.
We did this only as a first step towards our own concepts.
It served mainly to avoid pitfalls when differentiating knowledge transfer from decision making.
We still applied constant comparison to make sure these concepts really fit our purpose.  

When we followed our research interest, we quickly recognized that the
process problems often involved an
\emph{Unwillingness to Cooperate}, which in turn occurred in the context
of knowledge transfer activities.

\subsubsection{Finding the Main Concepts}

At this point, we had the sudden insight that the broken collaboration was likely a case
of a dysfunctional Teacher/Student setting.
These would be roles in the sense of \cite{SalPreZie13}.
So we went to work out these roles in the form of facets as described by \cite{SalPreZie13}.

This suggested employing yet another reusable piece of pair programming conceptualization
from prior work: Zieris' \emph{Push} and \emph{Pull} concepts for describing the structure
of knowledge transfer episodes:
\begin{itemize}
	\item \emph{Pull}: \say{Propellor pursues an internal Knowledge Want by eliciting explanations from partner who understands Topic and delivers Target Content.} \cite[Section 9.2.1]{Zie20}
	\item \emph{Push}: \say{Propellor pursues an external Knowledge Want by providing explanations to her partner without being requested to do so.} \cite[Section 9.2.1]{Zie20}
\end{itemize}

Five of the six facets of the \emph{Teacher} are GTM-style
expectations (``theoretical sensitivity'') we had on teacher behavior
that turned out to be fulfilled in our data. 
The sixth arose from a statement in one of our interviews.

\subsection{Manners of Concept Definition}\label{sec:conceptdefs}

The results section will use a wealth of GTM concepts.
Some of these are new, others have been introduced in earlier works
(Zieris' dissertation~\cite{Zie20}, the Base Layer book~\cite{SalPre13}).
Some of the new ones will be defined by an entire sections, others by
only one sentence, still others only implicitly by a suitable name.
The space we spend for the definition is commensurable with the
concept's overall relevance within the article.

\section{Results}

\subsection{RQ Tpos: Defining the Teacher Role} \label{ResDefTeacherRole}

We define the \emph{Teacher} role based on six facets (Sections~\ref{Fac_PKT}–\ref{Fac_ATL}).

To be a \emph{Teacher}, one needs to show \say{enough} such behavior,
as we will discuss along the way.
In our examples, we will use the terms 
\emph{CTeach} (for candidate teacher) and
\emph{CStud} (for candidate student) to clarify who is who
without making a final decision with respect to their \emph{Teacher}-ness
or \emph{Student}-ness.
Note that although each facet is in principle a single concept,
for simplicity we do not provide a closed definition of each (which
would often be hard to understand).
Rather, we provide a rough initial definition and then elaborate it by a
discussion that will often involve operationalization methods.

\subsubsection{Facet: Prioritizing Knowledge Transfer} \label{Fac_PKT}

The will to teach.
The attitude(!) of valuing successful knowledge transfer higher than
other outcomes of the session, in particular tangible work results.
This is the most important facet.  
If it is not present, consider the \emph{Teacher} role as not filled.  
\emph{Prioritizing Knowledge Transfer} is an intention,
thus not directly observable.
 
We operationalize it by differentiating \emph{Push} knowledge transfer
episodes based on
their purpose in the pair programming session:
Does the transferred knowledge directly impact the artifact being worked on? 
If so, we call it \emph{Mandatory Knowledge Transfer}.  
See Example~\ref{PA3_assertSame2sub}.
All other cases we call \emph{Optional Knowledge Transfer}.  
We further differentiate this into:
\begin{itemize} 
	\item \emph{Supplementary Explanation}, where knowledge is
      transferred that is related to the content of the session,
	\item \emph{Extra Explanation}, where the knowledge is not
      directly related to the content, and
	\item \emph{Follow-up Explanation}, where knowledge is transferred
      that explains a past decision. 
	 The same explanation could have been \emph{Mandatory Knowledge
     Transfer} if it had happened earlier in the session.
\end{itemize}
Such (observable) \emph{Optional Knowledge Transfer} suggests the presence
of the (only indirectly observable) facet
\emph{Prioritizing Knowledge Transfer}. 
See Examples~\ref{PA4_Getter_Setter} and~\ref{PA3_Tab}.

\begin{exmpl}[PA4 - Follow-up and Supplementary Explanation] \label{PA4_Getter_Setter}
	Driver P3 and \emph{CTeach} observer P1 decide to rename an attribute.  
	P3 starts to do so manually.  
	He consequently also changes the get- and set-methods in the corresponding class.  
	This exchange happens after he is done. 
	
	\DpartnerBC{P1}{The alternative would have been (!!...!!)\footnote{(!!...!!) symbolizes an interruption by the partner.}} \DCode{follow-up\_expl}
	
	\DpartnerA{P3}{Refactor.}
	
	\DpartnerBC{P1}{Exactly. To rename the member. Then the IDE would have done everything for you.} \DCode{follow-up\_expl}
	
	\DpartnerA{P3}{Does it even change the get- and set-methods?}
	
	\DpartnerBC{P1}{Yes, it even goes so far that if you rename the class, it renames the name of the class and possibly even the namespaces to which it belongs.} \DCode{suppl\_expl}
	
	\DpartnerAC{P3}{Nice.} \DCode{acknowledge\_helpful}
	
	P1's first \emph{Push} is labeled as \emph{Follow-up Explanation} because it happens after P3 is done.  
	P1's third remark is a \emph{Supplementary Explanation}.  
	Note that although P1 is initially answering P3's question (\emph{Pull}),  
	he continues to provide additional information, although a simple \say{yes} would have been sufficient.  
	It is a \emph{Pull} answer, followed by a \emph{Push}.
\end{exmpl}

\begin{exmpl}[PA3 - Extra Explanation] \label{PA3_Tab}
	\emph{CStud} P3 renames a test class using the IDE using an awkward, 
	inconvenient key sequence. 
	
	\DpartnerBC{P1}{If you use Tab, it completes it for you without you having to remove the test. So Tab always means ‘complete this’ and replaces what is underneath it} \DCode{extra\_expl}
	\DpartnerAC{P3}{Ah, cool.} \DCode{acknowledge\_helpful}
	
	\emph{CTeach} P1's remark is not concerned with the product they are working on. 
	It is a general piece of advice on how to better use the IDE and barely
	relevant for the progress within the present session. 
	That is why it is coded as \emph{Extra Explanation}.
\end{exmpl}

\subsubsection{Facet: Understanding Knowledge Needs} \label{Fac_ESoK}

Knowing what to teach.
Being a (good) \emph{Teacher} requires being aware of your partner's standard of knowledge. 
Striving for this awareness constitutes \emph{Understanding Knowledge Needs},
the second mandatory facet.
The Base Layer~\cite{SalPre13} provides the concept \emph{ask\_standard of knowledge} for direct questions, but there are indirect approaches as well.

The operationalization is based on two auxiliary concepts:
We annotate \emph{Recognize Unnoticed Knowledge Gap} when a
\emph{CTeach} recognizes the \emph{CStud} is doing something wrongly
or inefficiently and explains a better alternative.  
Usually, this is coded with \emph{challenge\_design} (or \emph{challenge\_step}) followed by \emph{explain\_knowledge}.  
Consider again Example~\ref{PA4_Getter_Setter}: 
\emph{CTeach} P1's first remark also qualifies as \emph{Recognize Unnoticed Knowledge Gap}.

Besides \emph{Recognize Unnoticed Knowledge Gap}, we also use the concept of \emph{Recognize Semi-Aware Knowledge Gap}.  
Here, the \emph{CStud} shows uncertainty by hesitant behavior, prolonged pauses, interrupting themselves, thinking aloud, or postponing a step.  
Again, the \emph{CTeach} reacts in a helpful way, typically by proposing and explaining a solution.  

\begin{exmpl}[PA3 - Recognize Semi-Aware Knowledge Gap] \label{PA3_expectedArgument}
	Driver \emph{CStud} P3 uses a function and seemingly does not know which argument to provide.  
	
	\DpartnerAC{P3}{\DDescription{Thinking aloud} The question now, of course, is what my (...) \DDescription{Writes \texttt{Null}}} \DCode{implicit\_uncertainty}
	
	\DpartnerBC{P1}{You now have to iterate over the overall result to the same extent as over your expected result, which of course makes it easier if you do a key-value operation. Entry is now only the value, you can also get key and value. } \DCode{Push}
	
	\DpartnerAC{P3}{\DDescription{Follows P1's suggestion.}} \DCode{implicit\_acknowledge}
	
	P3 is showing four different symptoms of uncertainty here: thinking aloud, interrupting himself, prolonged pause, and postponing the problem by writing \texttt{Null}. \emph{CTeach} P1 steps in in a helpful way.
	
\end{exmpl} 

Note that any \emph{CTeach} reaction to a \emph{CStud}'s \emph{direct} expression of uncertainty (for example by saying \say{I do not get it}, 
which would be annotated with \emph{explain\_standard of knowledge} \cite{SalPre13}) cannot trigger this facet.
Rather, this would be a \emph{Pull} by the \emph{CStud} and not
a proactive \emph{Push} by the \emph{CTeach}.

\subsubsection{Facet: Understanding Knowledge Want} \label{Fac_EKW}

Knowing when to teach.
Not every knowledge transfer is welcomed by the partner.  
A good \emph{Teacher} determines whether the partner is interested in the
knowledge the \emph{Teacher} could provide.
The most obvious way of doing so is by asking directly (annotated as
\emph{ask\_knowledge want}, which is not a Base Layer code).
See Example~\ref{DA2_classLoading}.

\begin{exmpl}[DA2 - Asking Knowledge Want] \label{DA2_classLoading}
	The software system used by D3 and \emph{CTeach} D4 is based on OSGi bundles,  
	a framework for developing and managing modular Java applications.   
	
	\DpartnerAC{D4}{Are you familiar with OSGi class loading and the like?} 
	
	\DCode{ask\_standard of knowledge}  
	
	\DpartnerB{D3}{Class, what?}
	
	\DpartnerA{D4}{Classloading}
	
	\DpartnerB{D3}{Not really. No.}
	
	\DpartnerAC{D4}{So, um.. Shall I explain briefly?} 	
	\DCode{ask\_knowledge want}  	
	
	\DpartnerB{D3}{Yes, of course!}
	
	Here we see how D4 first elicits the standard of knowledge of the \emph{CStud} by an \emph{ask\_standard of knowledge}, followed by asking for the knowledge want of his partner.
\end{exmpl} 

Such explicit behavior is infrequent in our data, but most
\emph{Pushes} are accepted without negative reactions (except in PA3).  
This means that either the \emph{CTeach} correctly senses the
knowledge wants of their partner or the \emph{CStuds} quietly accept 
\emph{Pushes} even if they are unwanted, which we find unlikely.
Rather we expect that partners who work together frequently develop
an intuition for each other's knowledge wants. 
Seeing such behavior by a \emph{CTeach} is a clue to good
\emph{Teacher} behavior, but this facet is optional.

\subsubsection{Facet: Pushing Helpful and Wanted Knowledge} \label{Fac_HP}  

Appropriate active teaching.
This facet is the consequence of the last two (see Sections~\ref{Fac_ESoK} and~\ref{Fac_EKW}).  
Only if a \emph{CTeach} pays attention to the knowledge needs and wants of the \emph{CStud}
are they able to \emph{Push Helpful and Wanted Knowledge}.  
Again, this is a sign of a good \emph{Teacher} but not mandatory.
 
But what is helpful and wanted knowledge? How do we recognize it?  
We deduce this based on the \emph{CStud}’s reaction.  
If they positively acknowledge the knowledge transfer,
we annotate it as \emph{Helpful and Wanted Push}.
The acknowledgment can be explicit
(see Example~\ref{PA3_assertSame2sub}, and again Examples~\ref{PA4_Getter_Setter} and~\ref{PA3_Tab})  
or implicit by incorporating the knowledge
(see again Example~\ref{PA3_expectedArgument}).

\begin{exmpl}[PA3 - Acknowledge Helpfulness] \label{PA3_assertSame2sub}
	\DpartnerA{P3}{\DDescription{Starts writing a for loop with \lstinline[language=PHP]|foreach ($entry in $result)|. This is underlined in red by the IDE.} Uhh, PHP does not have that? \DDescription{Smiles embarrassedly.}} 
	
	\noindent [\emph{CTeach} P1 explains the correct notation in PHP. P3 implements the for-loop where each element in the result of the serialize function is compared to its corresponding part in the map using \mbox{\texttt{assertSame}}. P3 reruns the test. It fails.] 	
	\DCode{letting\_mistakes\_happen} 
	
	\smallskip
	
	\DpartnerBC{P1}{Now we have a check for data types. However, we do not yet have the following checks. The question now is whether we want them or not. You are now iterating over your result. Suppose your result is empty. Then your test is true. Because you start iterating on the result.} \DCode{mandatory\_knowledge}
	
	\DpartnerAC{P3}{True, good point. \DDescription{Starts typing}} 	
	\DCodeB{acknowledge\_helpful}	
	
\end{exmpl}

Note that this facet is not based on actions of the \emph{CTeach} at all, 
but solely on the reactions of the \emph{CStud}.

\subsubsection{Facet: Providing on Pull} \label{Fac_PoP} 

Appropriate reactive teaching.
Besides proactively \emph{Pushing Knowledge}, a \emph{Teacher} will also
react helpfully to \emph{Pulls}.
Helpfulness is determined as in facet
\emph{Pushing Helpful and Wanted Knowledge} (\ref{Fac_HP}).

\begin{exmpl}[CA5 - Providing on Pull] \label{CA5_Warnings}
	\emph{CTeach} C3 and C4 are currently working on error messages that appear when the user creates a selection containing nothing. 
	
	\DpartnerB{C3}{\DDescription{Changes error message.}}
	
	\DpartnerAC{C4}{It depends on whether the geometry capture is still in the same condition, doesn't it?} \DCode{Pull}
	
	\DpartnerB{C3}{Yes, the \texttt{false} ensures this. So you just get out again and the dialogue is not closed.}
	
	\noindent [Dialogue has been omitted here to shorten the example.] \smallskip
	
	\DpartnerAC{C4}{But how do you now get to the other part?} \DCode{Pull}
	
	\DpartnerB{C3}{Erm, that's simply an intersection.}
	
	\DpartnerAC{C4}{Mm-hm. Intersection. Yes, ok, thanks.} 
	
	\DCode{acknowledge\_helpful}
	
	Here we see two \emph{Pulls} by C4 to which C3 responds by explaining. After the second explanation C4 acknowledges the helpfulness.
\end{exmpl}

\subsubsection{Facet: Allocating Time for Learning} \label{Fac_ATL}

Patience for teaching.
A \emph{Teacher} may emphasize learning even more than when
\emph{Prioritizing Knowledge Transfer} (see Section~\ref{Fac_PKT})
by \emph{Letting the CStud drive},
in extreme cases even including \emph{Letting mistakes happen}.
Both of these behaviors will slow down the session progress in terms
of practical work outputs.

\emph{Letting the CStud drive} is present in most of our examples:
\ref{PA4_Getter_Setter}, \ref{PA3_Tab}, \ref{PA3_expectedArgument}, \ref{PA3_assertSame2sub}, \ref{PA3_MagicNumber}, \ref{PA3_AssertSame}, \ref{PA3_NullObject}, \ref{PA3_runTests}, \ref{PA3_CodeStyle}.

\emph{Letting the CStud drive} does not always mean the 
\emph{CStud} actually will drive, as shown in the following example:

\begin{exmpl}[DA2 - Proposed Driver-Observer Switch] \label{DA2_PropSwitch}
	D3 and D4 asked a third developer, D6, for help approx. 30 minutes before this example. Since then, D4 was in control of the IDE because - as D3 mentioned - he understood better what D6 explained. Since then, D3 had trouble following along. 
	
	\DpartnerAC{D4}{Uh, do you want to take over again?} \DCode{hand\_control\_over}
	
	\DpartnerBC{D3}{I think you're more involved in the whole thing. I'm there... For me, it's already the top edge of knowledge. \DDescription{Laughing.}} 
	
	\DCode{reject\_control} 
	
	\DpartnerAC{D4}{\DDescription{Laughing.}} 
\end{exmpl} 

We have also already seen an example of the most extreme shape
of \emph{Allocating Time for Learning}, namely
\emph{Letting mistakes happen},
in Example~\ref{PA3_assertSame2sub}:
P1 presumably realized early on that P3's implementation of the for-loop is lacking but did not mention it until P3 was finished.  
We believe this is an instance of
\emph{Letting mistakes happen}.
In general, however, it is difficult to recognize this behavior.
Indeed, the concept had not occurred to us until an interviewee mentioned it:

\Qby{I5}{QRzDCLO}{The 
	other thing is when someone starts doing something in one direction  
	and you realize it's going to be a dead end.  
	You can -- and I've done this a few times, sometimes it works very well --  
	you can just let them go and say: \say{Yes, go for it}.  
	Then he'll see that it doesn't work or he'll know afterwards. [...]  
	That's a really good learning path.  
	If you have the time, that's good.  
	And of course you have to take yourself back.}

Note that Backseat Driving \cite{JonFle13}, i.e.,
directing a driver with low-level commands, is not
\emph{Letting the CStud drive},
because it creates essentially no opportunity for learning.

\subsection{RQ Spos: Defining the Student Role} \label{ResStudentRole}

Corresponding to the \emph{Teacher} role, we now define the \emph{Student} role
with two facets, both of them mandatory. 
If one partner is acting according to the \emph{Teacher} role,
the other will not necessarily take up the Student role.
A \emph{Student}, however, cannot exist without a
\emph{Teacher}.

\subsubsection{Facet: Providing a Back Channel} \label{Fac_Stu_PBC}

Supporting the teacher.
Once a knowledge transfer has happened where the
\emph{CStud} received knowledge from the \emph{CTeach},
\emph{Providing a Back Channel} means to signal how well the knowledge
was understood (\emph{explain\_standard of knowledge} \cite{SalPre13}), 
whether positively 
(e.g., \say{I got that}; see Example~\ref{DA2_AnonClass}) 
or negatively
(e.g., \say{I'm confused.}; see Example~\ref{CA5_SolutionDesign}). 

\begin{exmpl}[DA2 - Signaling Understanding] \label{DA2_AnonClass}
	\emph{CTeach} D4 is currently editing the line: \\ \texttt{action.execute(AbstractList.this.getPreview())} 
	
	\DpartnerA{D4}{\DDescription{Talking to himself} Why is \texttt{AbstractList} still in front of it? \DDescription{Deletes \texttt{AbstractList}. The IDE now highlights the line in red.} No. \DDescription{Undoes his action.} Oh, ok, ah. Because we're in the anonymous class.}
	
	\DpartnerBC{D3}{In an anonymous class? What is an anonymous class?} 
	
	\DCode{Pull}
	
	\DpartnerA{D4}{Erm, as soon as you implement a \texttt{SelectionAdapter} (!!...!!)}
	
	\DpartnerBC{D3}{Oh yes!} \DCode{signal\_understanding}
	
	\DpartnerA{D4}{here, for example, it is an anonymous class because the class has no name.}
	
	\DpartnerBC{D3}{Right. Right. Yes.} \DCodeB{signal\_understanding}
	
\end{exmpl} 

\begin{exmpl}[CA5 - Signaling Non-Understanding] \label{CA5_SolutionDesign}
	\emph{CTeach} C3 and C4 discuss the solution design, specifically how they should reuse parts of an existing method for a new feature.
	
	\DpartnerB{C3}{Mhm. I'm always torn about that. So I would almost tend to cut it out as it is in the first step and remove the duplications in the second step.}
	
	\DpartnerA{C4}{\DDescription{No reaction, C4 continues to look at the screen.}}
	
	\DpartnerB{C3}{But it doesn't really matter.}
	
	\DpartnerAC{C4}{I didn't understand that.} \DCode{signal\_non-understanding}
	
	\DpartnerB{C3}{I see two possibilities: either we pull what we think we need out of here into a method or reuse [the method] somewhere. [...]}
	
	\noindent [The exchange continues several more steps.] \smallskip
	
	As C3 and C4 discuss the solution design here, we categorize it as \emph{Mandatory Knowledge Transfer}, not actually as teaching. 
\end{exmpl} 

Asking follow-up questions is also a way of \emph{Providing a Back Channel}:

\begin{exmpl}[DA2 - Follow-up Questions] \label{DA2_Toolbar}
	\emph{CTeach} D3 and D4 are looking at the GUI. Their task is to add the same toolbar to multiple views.    
	
	\DpartnerAC{D4}{And these are now, uhh, is that a toolbar or coolbar or something that contains your own widgets in it, or what is that?} \DCode{Pull}
	
	\DpartnerB{D3}{Erm, I suppose so. I can't say for sure either, because I haven't done it yet either. Um, we'll just have to take a look at how it was done in [other part of GUI], for example.}
	
	\DpartnerAC{D4}{But, um, shouldn't the aim be that you always have these things here?} \DCode{ask\_follow-up}
	
	\DpartnerB{D3}{Yes, exactly.}
	
	\DpartnerAC{D4}{But the calendar (!!...!!)} \DCode{ask\_follow-up}
	
	\DpartnerB{D3}{Yes, what I told you earlier was rubbish.}
	
	\DpartnerAC{D4}{Yes, then that should go up there too, right?} 
	
	\DCode{ask\_follow-up}
	
	\DpartnerB{D3}{We won't put it here in this narrow bar, but up there.}
	
	We see how D4 gradually increases his understanding of the task by asking further questions after each explanation.
\end{exmpl}

\subsubsection{Facet: Expressing Knowledge Wants} \label{Fac_Stu_KWN}

Asking the teacher.
\emph{Expressing Knowledge Wants} is a \emph{Pull} in Zieris'
terminology. 
See again Examples~\ref{CA5_Warnings}, \ref{DA2_AnonClass}, and
\ref{DA2_Toolbar} for illustration.

\subsection{RQ Tneg: How to Fail as Teacher} \label{Res_Fail_T}

The success of knowledge transfer is highly variable.
Therefore, it is useful to execute the \emph{Teacher} and \emph{Student}
roles neatly.
Here, we describe five behavioral anti-patterns that can break the
\emph{Teacher} and \emph{Student} dynamic such that knowledge transfer fails
or is greatly diminished, four on the \emph{Teacher} side and
one on the \emph{Student} side.

\subsubsection{Anti-Pattern: Pushing Redundant Knowledge} \label{sec_fail_redundant}

Not knowing what not to teach.
Here, the Teacher fails at \emph{Understanding the Knowledge Need}.
This can lead to \emph{Pushing Redundant Knowledge}, where knowledge is provided that the partner already possesses. 
The following example illustrates the consequences; it is the most severe case we witnessed. Look for cases of \emph{push\_redundant}:

\begin{exmpl}[PA3 - Pushing Redundant Knowledge] \label{PA3_MagicNumber}
	Previous to this exchange, \emph{CTeach} P1 spotted a \say{magic number} ($0.01$) in the source code, 
	i.e., a constant value that is used in several places. 
	The pair decided to create the constant 
	OFFSET\_PER\-CENTAGE
	for this value. 
	After doing so, P3 runs the corresponding test.\footnote{Transcript by Zieris \cite[Section 9.6.3]{Zie20}.}
	
	\DpartnerBC{P1}{It’s important to make clear that the last two $0.01$ have no relationship. Because they might have no relationship and someone comes along and says \say{Oh, look, it says $00.1$ (!!...!!)}} 
	
	\DCode{push\_redundant}
	
	\DpartnerA{P3}{Which last two?}
	
	\DpartnerBC{P1}{The last two in lines 31 and 32, for example. Assuming the two numbers would have no relation and someone who only sees the implementation with raw numbers thinks \say{Oh, there is a relation, I’ll introduce a constant}. And then another comes along and introduces	it everywhere. Now all have the same relation. Now you know that they should explicitly be converted this way} \DCode{push\_redundant}
	
	\DpartnerA{P3}{If it’s the same relation, you can treat them as such. You adapt it the moment it changes}
	
	\DpartnerBC{P1}{Yes, but the one seeing the code doesn't know when you have only raw numbers, with	the same values. What about 3660?} 
	
	\DCode{push\_redundant}
	
	\DpartnerA{P3}{When did we ever have 3660 as a percentage?}
	
	\DpartnerBC{P1}{Or 3600! With 3600 it’s an example. That’s the conversion from hours to minutes, but also from seconds to minutes. Depending on the context two identical number can mean two completely different things.} \DCode{push\_redundant}
	
	\DpartnerA{P3}{But applied to our case this has no relevance.}
	
	\DpartnerBC{P1}{Yes, it has. Because it is a Magic Number, and Magic Number means (!!...!!)} \DCode{push\_redundant}
	
	\DpartnerA{P3}{But it is no longer \say{magic}. We just named it.}
	
	\DpartnerBC{P1}{\DDescription{Annoyed} Yes, we named it because it now creates a relation between these individual	numbers. Before, it was not clear (!!...!!)} 
	
	\DCode{push\_redundant}
	
	\DpartnerA{P3}{I don’t understand what you want, right now.}
	
	\DpartnerB{P1}{I wanted to explain why we are doing this (!!...!!)}
	
	\DpartnerA{P3}{\DDescription{Annoyed} I got that.}
	
	\DpartnerB{P1}{Good. It’s alright then.}
	
	\DpartnerA{P3}{\DDescription{Nervous laughter} I tried to understand what you still wanted to change.}
	
	\DpartnerB{P1}{Nothing. I didn't want to change anything.}
	
	\DpartnerA{P3}{\DDescription{Relieved} Ok.}
	
	\DpartnerB{P1}{I only want to clarify that it’s important to (!!...!!)} 
	
	
	\DpartnerA{P3}{\DDescription{Annoyed} Got it.}
	
	\DpartnerBC{P1}{make the relation with this renaming.} \DCode{push\_redundant}
	
	\DpartnerA{P3}{\DDescription{Annoyed. Stares at the screen} So.}
	
	\DpartnerBC{P1}{Not only to rename the variable.} \DCode{push\_redundant}
	
	\DpartnerA{P3}{\DDescription{Annoyed} It's ok.}
	
	\emph{Pushing Redundant Knowledge} can lead to frustration. 
	Multiple times do we see an annoyed P3 interrupting P1. 
\end{exmpl}

One of our senior interviewees, who has experience in taking up the \emph{Teacher} role, confirmed this anti-pattern directly:
\Qby{I5}{QRzDKTAnnoy1}{I can go on and on, and of course that annoys
  people when they already know things [or] that has nothing to do
  with the topic.}
Besides \emph{Pushing Redundant Knowledge}, this quote also addresses
\emph{Pushing Unwanted Knowledge}:

\subsubsection{Anti-Pattern: Pushing Unwanted Knowledge} \label{sec_fail_unwanted}

Not knowing what to teach or when.
Sometimes, the \emph{CStud} does not appreciate receiving new knowledge. 
This can even be the case if the knowledge is important to the session. 
In such cases it is a sign that the \emph{Teacher/Student} dynamic is dysfunctional:

\begin{exmpl}[PA3 - Pushing Unwanted Knowledge] \label{PA3_AssertSame}
	\emph{CTeach} P1 and P3 are currently fixing a test case for a serializer function. 
	The test case must compare the result of this function to a predefined array (in PHP, arrays are maps). 
	The pair is struggling to decide whether to use 
	\texttt{assertSame}, \texttt{assertEquals}, or yet another function.
	
	\DpartnerBC{P1}{So now, of course, with \texttt{assertSame} you have the additional condition that the two share the same memory area, that they have the same memory} \DCodeTwo{distancing\_you}{Push}
	
	\DpartnerA{P3}{\DDescription{P3 searches for an alternative to \texttt{assertSame} using the IDE suggestions while P1 speaks.} That this is the identical object.} 
	
	\DpartnerB{P1}{Exactly. Um}
	
	\DpartnerAC{P3}{\DDescription{scrolls to \texttt{assertEquals}} We want equals, I suppose.} 
	
	\DCode{Pull}
	
	\DpartnerB{P1}{Equals would be (!!...!!)}
	
	\DpartnerAC{P3}{I don't know if this works with Arrays} 
	
	\DCode{express\_uncertainty}
	
	\DpartnerBC{P1}{Equals is better, you can also use it to compare arrays. Mmh in case of doubt, you have a dependency on the order of the individual items that they contain, which is also (!!...!!)} 
	
	\DCodeThree{overcomplicated\_expl}{lack\_of\_focus}{distancing\_you}
	
	\DpartnerAC{P3}{There is even maxDepth \DDescription{Points to the screen.}} 
	
	\DCode{lack\_of\_back-channel}
	
	\DpartnerB{P1}{Not necessarily.}
	
	\DpartnerA{P3}{So cool!}
	
	\DpartnerBC{P1}{Er, which is not necessarily that important. But if you still want to ensure the data types, in other words you want to say, okay this is really a string that is returned, this is really a float (!!...!!)} 
	
	\DCodeThree{unstructured\_expl}{distancing\_you}{Push}
	
	\DpartnerAC{P3}{Mhm. \DDescription{continues scrolling through the IDE suggestions.}} 
	
	\DCode{lack\_of\_back-channel}
	
	\DpartnerB{P1}{or an object (!!...!!)}
	
	\DpartnerAC{P3}{Look, there is (!!...!!)} \DCode{lack\_of\_back-channel}
	
	\DpartnerBC{P1}{You can also iterate.} 
	
	\DCodeTwo{lack\_of\_responsiveness}{distancing\_you}
	
	\DpartnerA{P3}{There is also json.}
	
	\DpartnerBC{P1}{You can also use that.} \DCodeTwo{distancing\_you}{lack\_of\_focus}
	
	\DpartnerA{P3}{\DDescription{Continues with scrolling though the IDE suggestions.}}
	
	\DpartnerBC{P1}{There are quite a lot of functions that I... I have never really  used during normal implementation. Sometimes (!!...!!)} 
	
	\DCodeThree{lack\_of\_focus}{lack\_of\_responsiveness}{Push}
	
	\DpartnerAC{P3}{So what did you say?}
	
	\DpartnerB{P1}{PhpStorm recommends something.}
	
	\DpartnerAC{P3}{What's better than Equals? What did you just want to say?} \DCode{Pull}
	
	\DpartnerB{P1}{So \texttt{assertEquals} would be one option}
	
	\DpartnerA{P3}{\DDescription{P3 selects \texttt{assertEquals} from the IDE suggestions.}}
	
	\DpartnerBC{P1}{But then of course you would have to use... uhh \texttt{assertSame}. Sorry, \texttt{assertSame}! But then you just iterate over all the values of the array. So you do a for each with an \texttt{assertSame}} 
	
	\DCode{distancing\_you}
	
	\DpartnerAC{P3}{Oh well.} \DCode{lack\_of\_back-channel}
	
	\DpartnerBC{P1}{This ensures that the key you expect is defined and that the data type you have there is really the same and not just, uh, convertible to this data type} \DCodeThree{unstructured\_expl}{distancing\_you}{Push}
	
	\DpartnerAC{P3}{Let's do Equals first... I would say. \DDescription{P3 implements a naive check, ignoring P1's proposal on looping over the elements.}} 
	
	\DCode{lack\_of\_back-channel}
	
	\DpartnerBC{P1}{In this case, you are also responsible for ensuring that the order is correct. This is always nasty with array comparisons.} 
	
	\DCodeThree{overcomplicated\_expl}{Push}{distancing\_you}
	
	\DpartnerAC{P3}{So.\DDescription{Talking more to himself and signaling he is done with this decision}} \DCode{lack\_of\_back-channel}
	
	P3 clearly is not interested in P1's explanations, which he demonstrates by interrupting P1, starting new topics, or not engaging content-wise with P1's explanations. 
	As a consequence, the pair ends up choosing \texttt{AssertEquals} although \texttt{AssertSame} would have been the better option and P1 even said so.    
\end{exmpl}

One of our interviewees offered this behavior when asked under which
circumstances pair programming would not work well:
\Qby{I2}{QpKTharmful}{For example, someone who wants to lecture me and
  knows everything better. 
  Then I'd probably get to the point where I'd say, come on, let's stop this.}

\subsubsection{Anti-Pattern: Ineffective or Wrong Explanation} \label{sec_fail_ineffect}

Actual inappropriate teaching.
Explaining complex things is difficult. 
An \emph{Ineffective Explanation} can be hard for the \emph{CStud} to follow
which can lead to frustration and bad decisions. 
Example~\ref{PA3_AssertSame} shows four different kinds of
\emph{Ineffective Explanations}:
\begin{enumerate}
\item \emph{Unstructured Explanation}: an explanation that lacks clarity.
  In Example~\ref{PA3_AssertSame}, the main argument for \texttt{assertSame} is
  that it takes the data types into account, but the explanation does
  not mention this directly. 
\item \emph{Overcomplicated Explanation}: contains so much information that 
  the \emph{CStud} cannot grasp the main point. 
  P1 talks about the order of the items, which even he himself finds
  \say{not necessarily that important}.
\item \emph{Lack of Focus}: an explanation that never gets finished
  because the pair's dialog switches to a different topic.
  P1 gets sidetracked by P3's random injections (for example, regarding JSON).
\item \emph{Lack of Responsiveness}: continuing with an explanation
  although the partner is recognizably not following.
  P1 ignores P3's (non-)reactions that suggest P3 to be not paying
  attention or not understanding the explanation. 
\end{enumerate}
In Example~\ref{PA3_AssertSame} the combination of all four types,
leads to an impressively ineffective explanation.
As a result, P3 makes the bad decision to use \texttt{assertEquals}.

Obviously, knowledge transfer success is threatened still more
if the explanations are (partly) false (\emph{incorrect\_expl}).
We briefly see this in Example~\ref{PA3_MagicNumber}, where P1 claims
that 3600 (rather than 60) is the conversion factor for hours to
minutes or minutes to seconds.
In that case, though, the negative effect on the session is minimal.
This is very different in the following case, again from PA3:

\begin{exmpl}[PA3 - Incorrect Explanation] \label{PA3_NullObject}
	\emph{CStud} P3 just finished writing a test case. 
	P1 explains why the test case is not complete.
	
	%
	%
	%

	\DpartnerBC{P1}{You implemented direct calls to the invoice-data-delivery in the serializer, to these objects that are attached to the invoice-shop, yes?} 
	\DCode{distancing\_you}
	
	\DpartnerA{P3}{Yes.}
	
	\DpartnerBC{P1}{Shall we take another quick look? Because it is possible that this object is empty. We should also take that into account in the test.} \DCodeTwo{incorrect\_expl}{Push}
	
	\DpartnerA{P3}{Yes. \DDescription{P3 opens the serializer file.}} 
	
	\DpartnerBC{P1}{Because this delivery thing only exists for stores that have invoice type 2. Invoice type 1 does not have this. These objects are only persisted -- which is also part of the first story -- are only persisted if the store has invoice type 2. Invoice type 1 is not persisted with it.} \DCodeTwo{incorrect\_expl}{Push}
	
	\DpartnerA{P3}{Mhm.}
	
	\DpartnerBC{P1}{Now the question is, is this function only called here for stores that have the Invoice Type 2 or also for 1. If for 1, you have to reckon with zero pointers here. Namely exactly where you have the (!!...!!)} \DCodeThree{incorrect\_expl}{distancing\_you}{Push}
	
	\DpartnerA{P3}{I mean, you say this doesn't even exist, for those who don't have it.}
	
	\DpartnerB{P1}{Oh okay, it's just}
	
	\DpartnerAC{P3}{Because you are providing a DataDelivery}
	
	\DCode{dwell\_on\_mistake}
	
	\DpartnerB{P1}{Oh, I didn't even see that at first, ok}
	
	\DpartnerAC{P3}{and you've just said that it doesn't exist for the others (!!...!!)} 
	
	\DCode{dwell\_on\_mistake}
	
	\DpartnerB{P1}{Yes, you are right.}
	
	\DpartnerAC{P3}{that means you couldn't call it at all.} 	
	\DCode{dwell\_on\_mistake}
	
	\DpartnerB{P1}{Yeah, I didn't see it just now, that's why. I thought you (!!...!!)}
	
	\DpartnerAC{P3}{We might have to take that into account in the controller, but not here.}	
	\DCode{dwell\_on\_mistake} 
	
    At this point, P1 has been getting on P3's nerves for 55 minutes (see Figure~\ref{Timeline_PA3}).
    Now he is getting an explanation wrong.
    P3 notices the mistake and blames P1 for it multiple times.
    This strong reaction only makes sense in the overall context 
    of the session and is a consequence of the already-strained 
    pair relationship which may even get damaged further as a
    result of this interaction.
\end{exmpl}

\subsubsection{Anti-Pattern: Patronizing Communication Style} \label{sec_fail_patronizing}

Unappreciative behavior.
The above anti-patterns were about the content of explanations.
The present one is about the manner of explaining.
In PA3, we noticed explanations with a 
\emph{Patronizing Tone}, 
\emph{Condescending Laugh}, and 
\emph{Distancing You-Phrasing}. 
For diagnosing any of these, we need to see fitting responses by the
\emph{CStud}, which indeed we found:
P3 variously exhibits
\emph{Annoyed Reactions}, 
\emph{Embarrassed Reactions}, and 
\emph{Uncooperative Behavior} (usually non-reactions);
see Examples~\ref{PA3_MagicNumber} and~\ref{PA3_runTests}.

Note that in both of the examples following below,
\emph{Togetherness} appears undamaged; the \emph{Process Fluency} is good.
But the partners' willingness to cooperate is visibly harmed.
Understanding this type of situation involves so many complex additional
concepts (the central one being the \emph{Power Gap}) that we have to
treat it in a separate article~\cite{NN26b-CHASE}.

\begin{exmpl}[PA3 - Patronizing Communication Style] \label{PA3_runTests}
	\emph{CStud} P3 uses a company-internal \texttt{run\_tests.sh} script to run test cases. 
	This script seems to be implemented in a naive way and lacks features that PHPUnit provides (such as \texttt{-{}-filter}, \texttt{-{}-group}). 
	
	\DpartnerA{P3}{\DDescription{Executes \texttt{run\_tests.sh} in the terminal.}}
	
	\DpartnerBC{P1}{Did you know that with \texttt{run\_tests} you are missing a lot of features that PHPUnit offers? \DDescription{Laughs}} 
	
	\DCodeThree{patronizing\_tone}{condescending\_laugh}{distancing\_you}
	
	\DpartnerAC{P3}{I just realized anyway that we could actually filter. That's the other thing, because now he's doing all of them.} 
	
	\DCode{defensive\_justification}
	
	\noindent [Dialogue has been omitted here to shorten the example.] \smallskip
	
	\DpartnerA{P3}{\DDescription{Runs PHPUnit and gets an error message.}}
	
	\DpartnerBC{P1}{It's the wrong way round. \DDescription{P1 is referring to the order of P3's terminal commands.} \DDescription{Laughs}} 	
	\DCode{condescending\_laugh}
	
	\DpartnerAC{P3}{\DDescription{Grins.}} \DCode{embarrassed\_reaction}
	
	\DpartnerBC{P1}{Well, \texttt{ls} goes before \texttt{phpunit},
      but you are calling the program, so all you want to do now is
      specify the configuration, \texttt{-c} and \texttt{app}.}
    
    \DCode{distancing\_you} 
	
	\DpartnerA{P3}{\DDescription{Writes \texttt{-c app}.}} 
	
	\DpartnerBC{P1}{Exactly and then you can do, um, \texttt{-{}-filter} if you only want to do a single test.} \DCode{distancing\_you}
	
	\DpartnerAC{P3}{Good, it works. \DDescription{Writes \texttt{-{}-filter} and adds the specific test case.}} \DCode{lack\_of\_back-channel}
	
	Here we see P3 hardly acknowledging several factually correct and relevant
    explanations from P1. This is uncooperative behavior.
\end{exmpl}

P1 also distances himself from P3 by using \emph{Distancing
You-Phrasing} throughout the session.
He rarely refers to the pair as a team, 
but instead phrases his sentences as if P3 were
solely responsible for the task and P1 were only advising. 
Consider again Examples~\ref{PA3_AssertSame}, \ref{PA3_NullObject}, and \ref{PA3_runTests} and pay attention to P1's phrasing. 
See also the following dialog:

\begin{exmpl}[PA3 - Distancing You Phrasing] \label{PA3_CodeStyle}
	This exchange happens at the beginning of the session. 
	\emph{CStud} P3, who is not too familiar with PHP, is implementing a new PHP method.
	
	\DpartnerBC{P1}{Uhh, while we're typing. Code formatting. It's a bit, we do it a bit differently in PHP. \DDescription{Laughs.} You remember that mixed style with the functions, brace on the new line and no space between the function name and the argument paren?} 
	
	\DCodeThree{patronizing\_tone}{condescending\_laugh}{distancing\_you} 
	
	\DpartnerAC{P3}{\DDescription{Applies the feedback to the code.} Yeees. \DDescription{Annoyed.}} 
	
	\DCode{annoyed\_reaction}
	
	Here, P1 distinguishes himself (and presumably all other
    \say{proper PHP developers} from P3 by the phrase 
    \say{we do it a bit differently}. P3 reacts in an annoyed manner.
\end{exmpl}

This distancing is further increased by multiple \emph{Unagreed Disengagements} by P1, 
to briefly talk to colleagues while P3 continues working on their task and through P1's \emph{Dismissive Posture}:
he sits behind P3 and is often laid back far in his chair, 
with his hands behind his head.
P3 reacts to all this in a negative way, mostly by keeping his own communication to a minimum (\emph{Uncooperative Behavior}).

\subsection{RQ Sneg: How to Fail as Student} \label{Res_Fail_S}

\subsubsection{Anti-Pattern: Not Providing a Sufficient Back Channel} \label{sec_fail_back}

Starving the teacher.
As we illustrated in sections
\ref{sec_fail_redundant}-\ref{sec_fail_ineffect}, the cause of failing
as a \emph{CTeach} is often that he or she knows too little about 
(or pays too little attention to) the standard of knowledge or
knowledge wants of the \emph{CStud}.
The \emph{CStud} can aggravate this problem by failing to
provide feedback via a \emph{Back Channel} (see \ref{Fac_Stu_PBC}) and 
by insufficiently \emph{Expressing Knowledge Wants} (see \ref{Fac_Stu_KWN}). 

The latter is hard to pinpoint and to codify. 
But the lack of \emph{Providing a Back Channel} becomes clear when
comparing different knowledge transfer episodes with one another.
Especially in Example~\ref{PA3_AssertSame}, we see many instances
where P3 does not provide feedback on P1's knowledge transfer
attempts. See also Example~\ref{PA3_runTests}.

\section{Limitations}\label{sec_lim}

We do not expect to have seen all facets that could sensibly be viewed
to be part of the \emph{Teacher} or \emph{Student} role, 
or all behaviors that can break those roles;
simply because those behaviors never occurred in our data.
This does not invalidate our findings in any way, but they need to
be viewed as potentially incomplete.
We do not strive for a complete Grounded Theory and thus do not need
Theoretical Saturation.

Any of the facets and behaviors we describe could have been given
a somewhat different name or shape. 
This is an unavoidable consequence of the interpretivist epistemological stance
\cite[Chapter 9]{Charmaz14}
that is the bedrock of qualitative analysis and also not something 
that threatens the validity of the findings as we framed them.

In the present article, we do not attempt to determine and conceptualize 
the reasons for the behaviors we describe. 
This is a tactical decision:
The present article is meant to be accessible for a practitioner audience.
An analysis of underlying forces would reduce its readability and
distract from the article's main message.

We have not yet evaluated the usefulness of our conceptualization
for producing improved  \emph{Teacher/Student} behavior.
Such an evaluation would ideally be done by extensive field observation
and is thus left to future work.

\section{Conclusions and Further Work}\label{sec_concl}

We explain two roles often taken by pair programming partners
when knowledge is asymmetric between them:
\emph{Teacher} and  \emph{Student}.
We explain the \emph{Teacher} in terms of 
six facets of effective behavior and warn of
four behaviors that destroy useful outcomes from the intention to teach.
We explain \emph{Student} in terms of
two facets of effective Student behavior and warn of
one behavior that greatly hampers teaching.

Pair programming always involves lots of knowledge transfer.
Being good at knowledge transfer is therefore a big component of pair
programming skill.
In many cases, the knowledge transfer will occur in a Teacher/Student
constellation, so that understanding what to do and not to do in
Teacher/Student situations is a first step towards strong pair
programming skill.
The second step is then turning the respective behaviors into a habit.

\subsection{Advice for pair programmers}

Our many examples illustrate just how big the difference in effect is
between constructive \emph{Teacher} and  \emph{Student} behavior
and their broken variants.
So if you want to be a successful pair programmer,
use the facets as checklists and learning goals
and use the negative behaviors as warnings (for yourself)
and alarm signs (when you encounter them from your partner).
If your pair partner exhibits a problematic behavior,
go to the meta-level and discuss the behavior as such.
Our article provides material that hopefully allows to convince 
your partner that your negative perception is to be taken seriously.

\subsection{Further work}

As an obvious next step, one should work out the conceptualization
into a full Grounded Theory of what is going on in the broken
\emph{Teacher/Student} constellations.
In fact, we have done a large part of this already;
the resulting Grounded Theory is written up in \cite{NN26b-CHASE}.

The habit-forming of someone attempting to learn good Teacher/Student
behavior is also likely non-trivial and methods for supporting it
should be investigated.

We recommend further research into other pair programmer roles as well.
\cite{SalPreZie13}, which explains a role definition framework,
also mentions several candidate roles.
We are sure further roles exist, if not as frequent as \emph{Teacher} 
and  \emph{Student}.
We believe that much like design patterns are highly effective for 
teaching program structures,
roles are a good means for explaining pair programming skill.

Finally, evaluating that (or even how much) this knowledge actually
helps pairs to be effective is a difficult but scientifically rewarding
task.


\section*{Acknowledgments}
We would like to thank all developers who participated in this study, either by allowing their pair programming sessions to be recorded or by lending us their time for interviews.

As English is not our native language, we used LLMs to help translate parts of the interviews and transcripts of the pair programming sessions. LLMs were also used to check spelling and grammar after writing the present article. 

Not a single sentence in the present article was written by an LLM. All code definitions and annotations are our own.

\bibliographystyle{ACM-Reference-Format}
\bibliography{special}

\end{document}